\documentclass[preprint,showpacs,preprintnumbers,amsmath,amssymb,superscriptaddress]{revtex4}
\usepackage{graphicx}
\begin{document}

\title{Strong-coupling \emph{d}-wave superconductivity in
PuCoGa$_5$ probed by point contact spectroscopy}

\author{D. Daghero}
\author{ M. Tortello}
\author{G.A. Ummarino}
\affiliation{Dipartimento di Fisica, Politecnico di
Torino, Corso Duca degli Abruzzi 24, 10129 Torino, Italy}

\author{J.-C. Griveau}
\author{E. Colineau}
\author{R. Eloirdi}
\affiliation{European Commission, Joint Research Centre, Institute
for Transuranium Elements, Postfach 2340, D-76125 Karlsruhe,
Germany}

\author{A. B. Shick}
\affiliation{European Commission, Joint Research Centre, Institute
for Transuranium Elements, Postfach 2340, D-76125 Karlsruhe,
Germany}
\affiliation{Institute of Physics, ASCR, Na Slovance 2, CZ-18221
Prague, Czech Republic}

\author{J. Kolorenc}
\affiliation{Institute of Physics, ASCR, Na Slovance 2, CZ-18221
Prague, Czech Republic}

\author{A. I. Lichtenstein}
\affiliation{University of Hamburg, Jungiusstrasse 9, 20355 Hamburg,
Germany}

\author{R. Caciuffo}
\affiliation{European Commission, Joint Research Centre, Institute
for Transuranium Elements, Postfach 2340, D-76125 Karlsruhe,
Germany}

\date{\today}
%\keywords{point contact spectroscopy, nonconventional superconductivity, heavy fermion superconductors, electronic structure calculations}
\maketitle

{\bf A century on from its discovery, a complete fundamental
understanding of superconductivity is still missing. Considerable
research efforts are currently devoted to elucidating mechanisms by
which pairs of electrons can bind together through the mediation of
a boson field different than the one associated to the vibrations of
a crystal lattice. PuCoGa$_5$, a $5f$-electron heavy-fermion
superconductor with a record critical temperature $T_c=18.5$ K, is
one of the many compounds for which the short-range, isotropic
attraction provided by simple electron-phonon coupling does not
appear as an adequate glue for electron pairing. Magnetic, or
virtual valence fluctuations may have an important role in the
stabilization of the superconducting ground state in PuCoGa$_5$, but
the specific nature of the coupling mechanism remains obscure. Here,
we report the results of point-contact spectroscopy measurements in
single crystals of PuCoGa$_5$. Andreev reflection structures are
clearly observed in the low-temperature spectra, and unambiguously
prove that the paired superconducting electrons have wavefunction
with the $d$-wave symmetry of a four-leaf clover. A
straightforward analysis of the spectra provide the amplitude of the
gap and its temperature dependence, $\Delta(T)$. We obtain
$\Delta(T \rightarrow 0)$ = 5.1 $\pm$ 0.3 meV and a gap ratio, $2\Delta/k_B T_c$ =
6.5 $\pm$ 0.3, indicating that the compound is in the regime of
strong electron-boson coupling. The gap value and its temperature
dependence can be well reproduced within the Eliashberg theory for
superconductivity if the spectral function of the mediating bosons
has a spin-fluctuations-like shape, with a peak energy of 6.5 meV.
Electronic structure calculations, combining the local density
approximation with an exact diagonalization of the Anderson impurity
model, provide a hint about the possible origin of the fluctuations.
}

PuCoGa$_5$, discovered in 2002, has the highest critical temperature
($T_c$ =18.5 K) among heavy-fermion superconductors \cite{sarrao02}.
Though the $\alpha$-activity and the toxicity of Pu make it
unsuitable for practical purposes, this material has being
challenging the researchers since its discovery. As a matter of
fact, it seems to elude any classification. The Sommerfeld
coefficient $\gamma$, as measured by specific-heat experiments, is
between 58 and 95 mJ/(mol K) \cite{sarrao02,bauer04}. The effective
mass of electrons is about $1/5$ of that measured for the
isostructural unconventional superconductor CeCoIn$_5$,
\cite{bauer04}, suggesting a smaller degree of electronic
correlation.

The magnetic properties of PuCoGa$_5$ and their possible
relationship with the superconducting pairing have been debated, mainly because of conflicting results about the
magnitude of the magnetic moment carried by the Pu ions. A Curie-Weiss susceptibility in the normal
state was initially observed \cite{sarrao02}, as expected for
fluctuating local $\mathrm{Pu}^{3+}$ moments. Successive $\mu$SR
studies showed no static, normal-state electronic magnetism in
PuCoGa$_5$ \cite{heffner06,morris06} and polarized neutron
diffraction experiments showed a small and temperature-independent microscopic
magnetization dominated by the orbital moment \cite{hiess08}.

The amplitude of the gap (or at least of the gap ratio
$2\Delta/k_B T_c$) has been estimated by different techniques. Using nuclear magnetic resonance (NMR) to measure the Knight shift in the superconducting state of a sample with  $T_c$=18.5 K, Curro \textit{et al.} obtain $2\Delta/k_B T_c$ = 8 ($\Delta \sim$ 6.3 meV) \cite{curro05}.

From the theoretical and computational point of view, PuCoGa$_5$ is
a real challenge. Various band structure calculations have been
reported
\cite{opahle03,maehira03,opahle04,tanaka04,shick05,pourovskii06,zhu11,shick11},
showing that the details of the Fermi surface (FS) and the value of
the local magnetic moment depend on the theoretical approach. In a
local density/generalized gradient approximations (LDA/GGA), the FS
is quasi-two dimensional \cite{opahle03,maehira03,shick05,zhu11},
with at least two nearly-cylindrical sheets, one hole-like around
the $\Gamma=(0,0,0)$ point of the reciprocal space $(k_x,k_y,k_z)$,
and one electron-like around the $M=(\pi,\pi,0)$ point. This bears a
similarity to what more recently observed in Fe-based
superconductors. The coupling mechanism that seems to account for
the high-$T_c$ superconductivity in iron pnictides (based on the
nesting between hole and electron FS sheets through a vector
associated with a peak in the spin susceptibility)  had been
proposed earlier for PuCoGa$_5$ \cite{tanaka04}. The FS remains
qualitatively unchanged with the addition of the Coulomb-$U$
interaction \cite{shick11}.

Several experimental facts
\cite{bauer04,bang04,curro05,boulet05,heffner06,ohishi07,jutier08}
suggest that in PuCoGa$_5$ the electrons in the pairs have a mutual
angular momentum $\ell$ = 2, corresponding to a superconducting
order parameter (OP) with $d_{x^{2}-y^{2}}$ symmetry, \textit{i. e.}
the gap in the single-particle excitation spectrum has a line of
nodes intersecting the Fermi surface. A direct proof of the gap
symmetry is, however, still missing. This is an important point, as
the symmetry of the OP is closely related to the pairing mechanism.
For instance, whilst isotropic electron-phonon attraction favors the
formation of zero-angular-momentum pairs, with a spherically
symmetric OP, $d$-wave symmetry is most easily realized if the
pairing interaction is repulsive at short-range and anisotropic at
larger distances, as the one provided by effective spin-spin
couplings on the border of antiferromagnetism \cite{monthoux07}.

A magnetic nature of the Cooper pairing mediator in PuCoGa$_5$ has
been discussed by several authors (see Ref. \cite{pfleiderer09} for
a recent review). Flint et al. \cite{flint08} considered virtual
valence fluctuations of the magnetic Pu configurations, creating
Kondo screening channels with different symmetry, and demonstrated
that in a lattice of magnetic ions, exchanging spin with conduction
electrons in two different channels, a condensate of composite pairs
between local moments and electrons is formed. These models must be
reconciled with the temperature-independent susceptibility observed
in the normal state \cite{hiess08}. A phononic mechanism in the
framework of the $d$-wave Eliashberg theory has been discussed in
Ref. \cite{jutier08}. Although able to reproduce a number of
experimental observations, this model requires an electron-phonon
(\textit{e-ph}) coupling constant $\lambda$ that is much higher than
the value experimentally deduced from the time relaxation of
photo-induced quasiparticles \cite{talbayev10} ($\lambda$ =
0.2-0.26).

%In this paper we report \emph{direct} spectroscopic measurements %of
%the gap in PuCoGa$_5$ single crystals by means of point-contact
%spectroscopy. We show that the experimental spectra
%unambiguously prove that the gap in PuCoGa$_5$ has nodes
%(i.e. it changes sign on the same FS sheet), and that its amplitude
%is $\Delta = 5.1 \pm 0.3$ meV for $T_c=18.5$K. This results in a
%strong-coupling gap ratio $2\Delta/k_B T_c \simeq 6.4 \pm 0.4$, %which
%is much higher than the BCS value 4.28 expected for a %weak-coupling
%$d$-wave superconductor. The gap amplitude
%scales with the $T_c$ of the crystals when the latter is
%suppressed by impurities or defects, so as to keep the gap ratio
%constant.

We performed point-contact Andreev-reflection spectroscopy (PCARS)
measurements in freshly annealed single crystals of $^{239}$PuCoGa$_5$,
grown by a flux method and  characterized by x-ray diffraction, electrical resistivity, magnetization and specific heat measurements. The point contacts were made between a fresh,
mirror-like surface of the crystal (just exposed by breaking the
sample) and a thin Au wire (about 10 $\mu$m in diameter).
%The
%resulting contacts are rather unstable both mechanically and
%electrically so that extreme cautions must be taken to avoid shocks
%(especially during cooling or heating).
The uneven broken surface on
which the contact is made prevents a fine control of the direction
of current injection with respect to the crystallographic axes. However, as shown below, this does not prevent
the unambiguous determination of the amplitude and symmetry of the OP.

\begin{figure}[htbp]
\includegraphics[width=.8\textwidth]{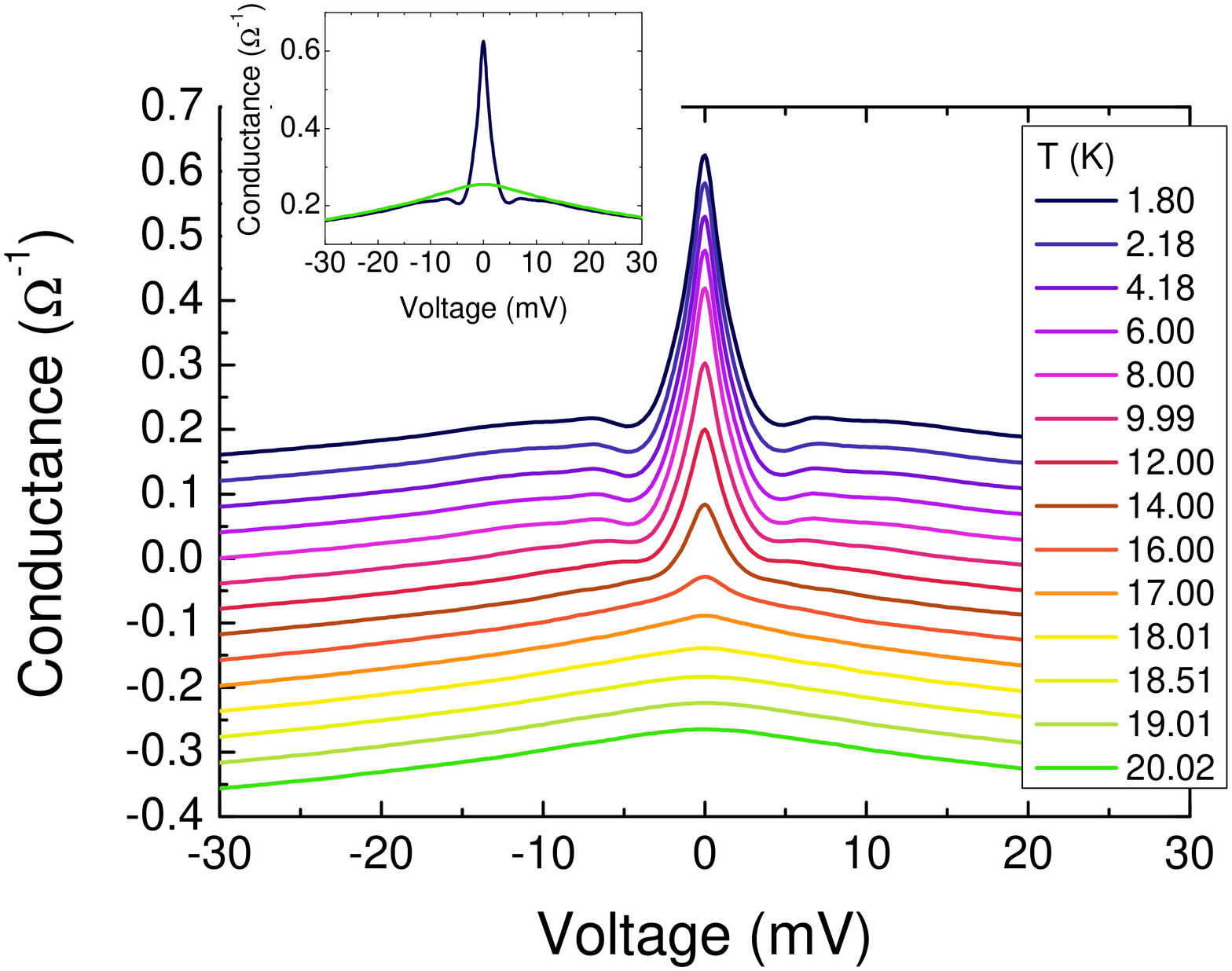}
\caption{Temperature dependence of the raw conductance curve of a
Au/PuCoGa$_5$ point contact with $R_N=6.2 \, \Omega$. All the curves
apart from the top one are vertically shifted for clarity. The inset
shows the curve at 1.8 K and the curve at 20.02 K (and thus in the
normal state) without any shift. The superposition of the tails is
one of the indicators that the contact is in the ballistic
regime.}\label{Tdep1}
\end{figure}

Fig. \ref{Tdep1} presents the temperature dependence of the raw
conductance curves of a point contact whose normal-state resistance
is 6.2 $\Omega$. All the curves but the lowest-temperature one are
shifted downward for clarity. The top curve is measured at $T=1.8$
K, that means about $T_c /10$. This ensures that the gap  extracted
from it is well representative of the gap for $T\rightarrow 0$. The
high-energy tails of the conductance curves are temperature
independent, as shown in the inset, where the curves measured at
$T=1.8$ K and in the normal state are reported without vertical
offset as an example. This demonstrates that the contact is in the
perfectly ballistic regime, with no contribution from Maxwell terms
in the contact resistance, i.e. there is no diffusion in the contact
region and the maximum excess energy with which the electrons are
injected into the superconductor is exactly $eV$. Meeting this
condition is essential for energy-resolved spectroscopy to be
possible. The absence of heating effects in the contact is also
witnessed by the coincidence between the \textit{Andreev critical
temperature} $T_c^A$ (i.e. the temperature at which the Andreev
signal disappears and the normal-state conductance is recovered,
here between 18.0 K and 18.5 K) and the bulk $T_c$, as determined
from resistivity measurements (T$_c$ = 18.1 K, see Fig. 6 in the
Supplementary Material).

It is worth noticing that the point-contact curves in PuCoGa$_5$ do
not show the strong asymmetry observed for CeCoIn$_5$ below the
temperature where a coherent heavy-fermion liquid develops~
\cite{park08}. The asymmetric conductance in CeCoIn$_5$ has been
explained in terms of a Fano resonance involving localized states
near the interface and itinerant heavy electrons in the
bulk~\cite{fogelstrom10}. The absence of a strong, temperature
dependent asymmetry in PuCoGa$_5$ may thus confirm that
superconductivity develops out of an incoherent metallic state
\cite{coleman06}.

In order to compare the experimental curves to the theoretical ones,
a normalization is required, i.e. a division by the conductance of
the same point contact when the superconductor is in the normal
state. Because of the extremely high value of the upper critical
field in PuCoGa$_5$, this cannot be achieved by applying a magnetic
field to suppress superconductivity at low temperature. However,
owing to the moderate $T_c$ and the negligible temperature
dependence of the normal state conductance (witnessed by the absence
of any change in shape of the tails of the conductance curves at
$|eV|>15$ meV) we can safely normalize the conductance curves at any $T<T_c^A$ by the normal state conductance curve measured at (or just above) $T_c^A$.

\begin{figure}[htbp]
\includegraphics[width=.6\textwidth]{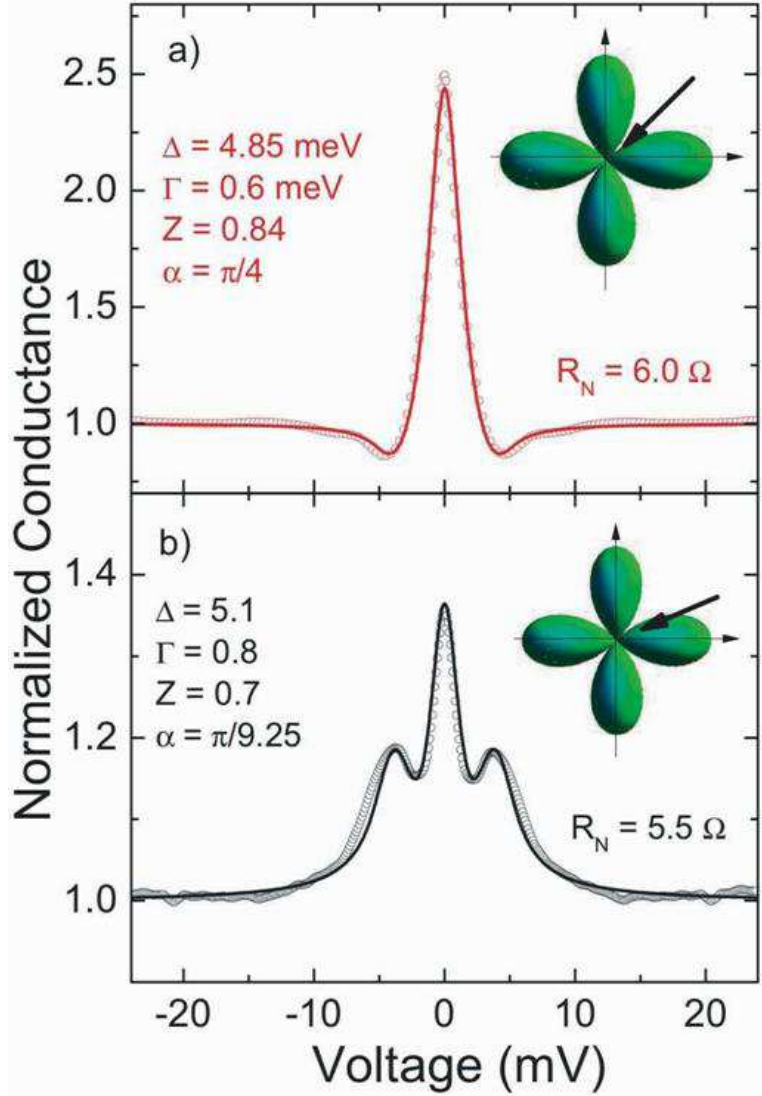}
\caption{Two examples of low-temperature conductance curves in
PuCoGa$_5$ after normalization (symbols) and the relevant 2D-BTK fit
assuming a $d$-wave gap (lines). The values of the parameters are
indicated in labels, as well as the normal-state resistance $R_{N}$
of the corresponding Au/PuCoGa$_5$ point contact. $\Delta$ is the
gap at $T$ = 0, $\Gamma$ is a spectral broadening parameter, $Z$
measures the barrier strength at the interface, and $\alpha$ is the
angle between the direction of current injection and the $k_{x}$
axis. The different shape of the two curves simply arises from the
different direction of the current injected with respect to the
lobes of the $d$-wave symmetry (insets). }\label{lowT}
\end{figure}

The result of this normalization is shown in Fig. \ref{lowT}
for two contacts made on different places of freshly-broken surfaces of the same
sample. The two contacts have the same $T_c^A$ but different
normal-state resistance $R_N$. The curve shown in Fig. \ref{lowT}a
presents a very clear zero-bias conductance peak (ZBCP), whose
amplitude (greater than 2) can only be explained by assuming a
nodal OP with a change of sign at the Fermi surface. Indeed, the shape of the conductance curve is similar to those shown for a dimensionless barrier strength $Z=1$ \cite{blonder82,kashiwaya96} in Fig. (1d)
of the Supplementary Material, and
clearly indicates that the OP has a $d$-wave symmetry.
The peak is ascribed to zero-energy Andreev bound states,
arising from the constructive interference between electron-like and
hole-like quasiparticles that feel order parameters with different
signs \cite{daghero10}.

Another important point to notice is that,
unlike in other heavy-fermion compounds like CeCoIn$_5$
\cite{fogelstrom10}, the Andreev signal is very high in the superconducting phase; the curve is
similar to the best ones ever observed in cuprates
\cite{deutscher05,wei98,biswas02,qazilbash03}. It thus seems that in
PuCoGa$_5$ there is no need to assume multichannel tunneling in the
presence of localized states near the interface, as instead is the case in CeCoIn$_5$ \cite{fogelstrom10}.

The conductance curve shown in Fig. \ref{lowT}b has a very different shape than the curve in Fig. \ref{lowT}a, reflecting a change from  $\alpha$ = $\pi/4$ to $\alpha$ = $\pi/8$ of the angle between the direction of the current injection (here in the $k_{x}$-$k_{y}$ plane) and the $k_{x}$ axis. As a matter of fact, the curve in Fig. \ref{lowT}b looks similar to those shown in Fig.~(1c) of the Supplementary Material, which have been obtained at T = 0 for intermediate values of $Z$ (0.5-0.7) and a
misorientation angle $\alpha=\pi/8$.

Solid lines in Fig. \ref{lowT} are the results of a quantitative spectral analysis using the 2-dimensional  Blonder-Tinkham-Klapwijk (2D-BTK) formalism \cite{blonder82,kashiwaya96,daghero10}. The parameters of the model are $\Delta$ (here
intended as the maximum amplitude of the gap), the barrier strength $Z$, the angle $\alpha$, and a
broadening parameter $\Gamma$ that must usually be
included in the model when fitting experimental curves. Here $\Gamma \ll \Delta$ so that it does not add any ambiguity to the
determination of the gap amplitude. The best-fitting values of these
parameters are indicated in the labels. The fit of the curves gives
very similar values of the gap, and this is certainly the most
important result, but also the barrier and broadening parameters are
very similar. This means that the very different shape of the two
curves almost completely depends on the angle $\alpha$ which is
different in the two cases, as shown pictorially in the insets.

\begin{figure}[htbp]
\includegraphics[width=.8\textwidth]{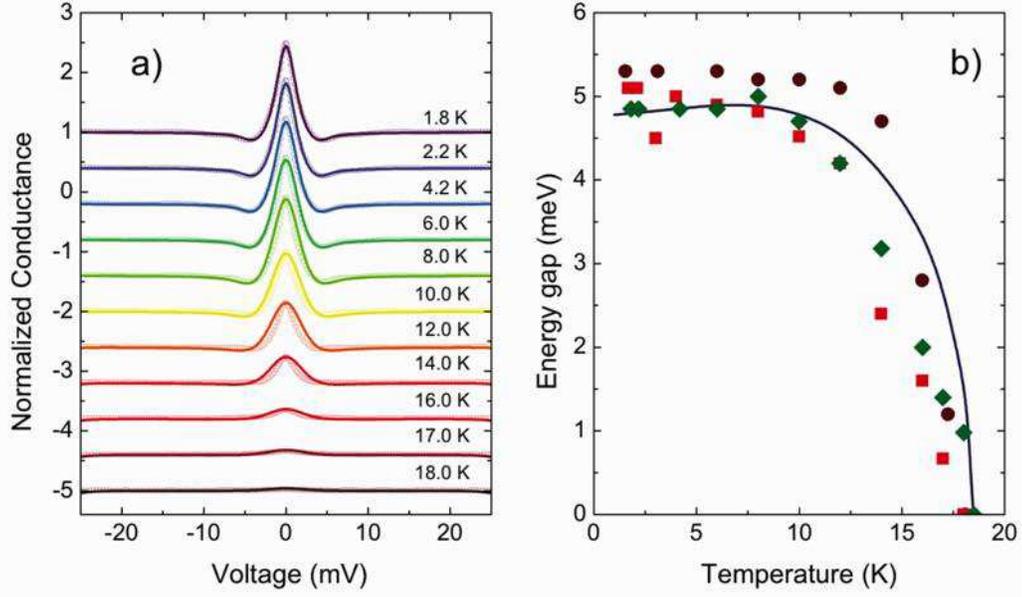}
\caption{(a) Temperature dependence of the normalized conductance
curves of a contact with $R_N$ = 6.0 $\Omega$ (symbols) and their
fit with the 2D-BTK model (lines). The curves are vertically offset
for clarity. (b) Temperature dependence of the gap extracted from
the fit of the conductance curves of three different contacts
(symbols). The solid line is the $\Delta(T)$ function calculated
within the Eliashberg theory by assuming a spin-fluctuation spectrum
peaked at $\Omega_0=6.5$ meV and a coupling constant $\lambda=2.37$.
}\label{gaps}
\end{figure}

Fig. \ref{gaps}a shows an example of how the normalized
conductance curves evolve on increasing temperature. The
experimental data (symbols) are compared to the 2D-BTK fit (solid
lines). The fit is rather good at any temperature; note that both
$Z$ and $\alpha$ are independent on T and were thus kept constant
for all curves. We also kept $\Gamma$ nearly constant so that the
only parameter that varies significantly with temperature is the gap
amplitude $\Delta$.

The temperature dependence of the gap as obtained from the fit of
three different sets of conductance curves (in three different
contacts) is shown in Fig. \ref{gaps}b. It is clear that the values
are rather well reproducible; at low temperature, the gap values
range between 4.85 and 5.30 meV, corresponding to a gap ratio
$2\Delta/k_B T_c = 6.2 - 6.7$. The vertical spread of gap values can
be used to evaluate \emph{a posteriori} the uncertainty on the gap
itself; this is obviously much greater than the uncertainty arising
from the fit of a single curve, which can be empirically determined
as the range of gap values that allow an acceptable (i.e. within
some confidence limit) fit of the conductance curve, when all the
other parameters are varied as well. The spread of gap values is
small at low temperature and maximum around 14 K. Although the
general trend of the gap seems to be compatible with a BCS-like
$\Delta(T)$ dependence (but of course, with a non-BCS gap ratio),
this uncertainty does not allow discussing it in detail.

The solid line in Fig. \ref{gaps}b is
\emph{not} a fit of the data, but is calculated within the
strong-coupling theory for superconductivity (known as Eliashberg
theory \cite{eliashberg60,carbotte90}) by assuming that the superconducting coupling is mediated by
spin fluctuations. In particular, we used the typical spin-fluctuation
spectrum with the form \cite{millis90}:
$\alpha^{2}F_{d}(\Omega)\propto\frac{\Omega
\Omega_{0}\vartheta(\Omega_{max}-\Omega)}{\Omega^{2}+\Omega_{0}^{2}}$
where $\Omega_{0}$ is the energy peak, $\Omega_{max}$ is a cut off
energy that we chose as being $\Omega_{max}=4 \Omega_{0}$
\cite{monthoux91}. The line in Fig. \ref{gaps}b
is obtained by taking $\Omega_0=6.5$ meV and using an
electron-boson coupling constant $\lambda= 2.37$.
Details of the calculations are given as Supplementary Material.
Assuming
physically reasonable values of the Coulomb pseudopotential $\mu^{*}$ ($0 \leq \mu^{*} \leq 0.2$) \cite{carbotte90}, the
observed $\Delta(T)$ curve pose a stringent limit on the
characteristic boson energy, namely $5.3 \leq \Omega_0 \leq 8$ meV. This energy range is compatible with that determined by recent NMR
measurements \cite{baek10} for the characteristic spin-fluctuation energy in PuCoGa$_{5}$, that is $4 \leq \Omega_{S}\leq 8$ meV.

As a test of the reliability and generality of the results
discussed so far, we also performed PCARS measurements in
$\mathrm{^{242}PuCoGa_5}$ crystals featuring a lower $T_c$ =14.5 K, due to the presence of Sb impurities (less than 1 \%). The experimental results (shown as Supplementary Material) have been analyzed following the procedure described above, using the \emph{same} spectral function and coupling constant as in Fig. \ref{gaps}b. We obtain a gap ratio $2\Delta/k_B T_c=6.2 \pm 0.4$, showing that $\Delta(T \rightarrow 0)$ scales with $T_c$.

A scenario where magnetic fluctuations are responsible for the
formation of the Cooper pairs in PuCoGa$_{5}$ must be reconciled with
the observed temperature-independent magnetic susceptibility
\cite{hiess08} that points to vanishing local moments at the Pu
sites. A plausible explanation is provided by electronic structure
calculations combining the local density approximation (LDA) with the
exact diagonalization (ED)~\cite{JindraED} of a discretized
single-impurity Anderson model~\cite{Hewson}. In this approach, the
band structure obtained by the relativistic version of the
full-potential linearized augmented plane wave method
(FP-LAPW)~\cite{shick01} is consistently extended to account for the
full structure of the $f$-orbital atomic multiplets and their
hybridization with the conduction bands~\cite{shick09}. Details on
this procedure are given in the Supplementary Material.

% With this approach, the structure of the
% full-atomic multiplet and the hybridization with the conduction band
% are treated on an equal footing with the band structure obtained by
% the relativistic version of the full-potential linearized augmented
% plane wave method (FP-LAPW)~\cite{shick01}.

\begin{figure}[htbp]
\includegraphics[width=.8\textwidth]{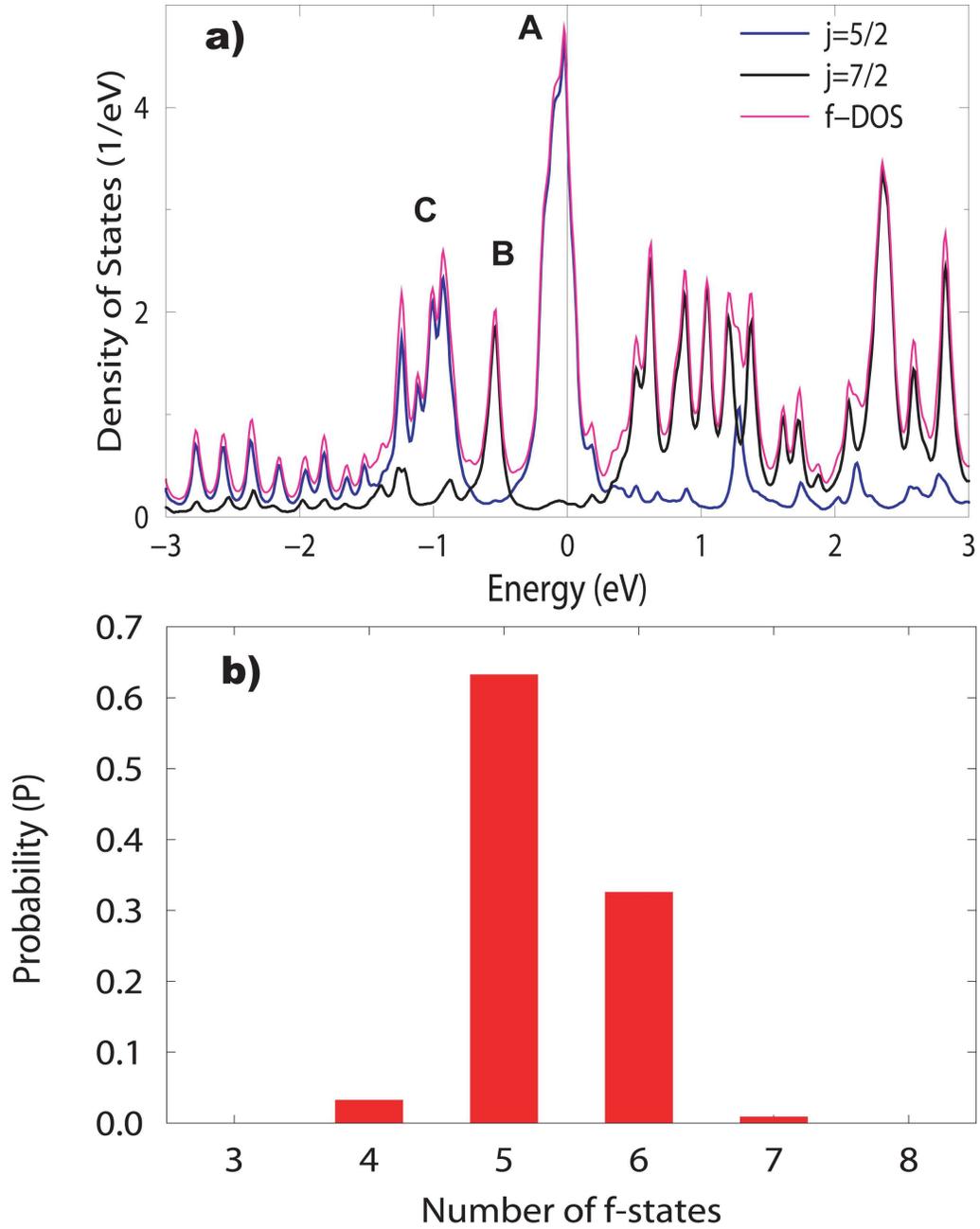}
\caption{(a) $f$-electron density of states (DOS) for the Pu atom in
PuCoGa$_5$: $j$ = 5/2, 7/2 projected, and total; (b) Valence
histogram obtained by the projection of the LDA+ED solution onto Pu
atomic eigenfunctions. }\label{mixval}
\end{figure}

The resulting $f$-orbital density of states (DOS) is shown in
Fig.~\ref{mixval}a. Below the Fermi energy $E_F$, the DOS exhibits a
three-peak structure that is typical for Pu and for a number of its
compounds. Peaks characterized by the total moment $j=5/2$ occur at
$E_F$ (peak labelled~A) and at about 1 eV below $E_F$ (peak~C). The
$j=7/2$ peak~B is located in between at approximately 0.5 eV below
$E_F$. The DOS is in a reasonably good agreement with the results of
the non-crossing approximation reported in Ref.~\cite{pezzoli11}.

Fig.~\ref{mixval}b shows the valence histogram calculated by
projecting the ground state of the impurity model $|\Omega\rangle$
onto the Pu atomic eigenstates $|m\rangle$ that correspond to an
integer $f$-shell occupation $n_m$. The plotted probabilities $P_m$
determine the $f$-orbital valence $\langle n_f \rangle = \sum_{m} P_m
n_m=5.3$. The highest obtained probability is $P_5=0.63$, followed by
$P_6=0.33$. In addition, there are small but non-zero probabilities
$P_4=0.03$ and $P_7=0.01$. This result indicates a mixed-valence
nature of the Pu $f$ shell that consists of a mixture of a magnetic
$f^5$ sextet and a non-magnetic $f^6$ singlet, which is similar to the
case of $\delta$-Pu~\cite{shim07}. This mixed (also referred to as
intermediate) valence is consistent with a large value of the
$\gamma$-coefficient, as typically observed in the mixed-valence
rare-earth-based materials. The same materials display a
temperature-independent magnetic susceptibility at low temperatures if
one of the involved atomic configurations is
non-magnetic~\cite{lawrence81}, which further strengthens the analogy
between the mixed-valence rare-earth compounds and the $f$-electron
physics in PuCoGa$_5$.

The Pu $f$ shell carries a non-vanishing average moment as it
fluctuates between the singlet and the sextet. The expectation values
$\langle \hat X^2 \rangle = X(X+1)$, where $X$ = $S$, $L$ and $J$,
calculated for the $f$ shell give $S^{(5f)}$ = 2.18, $L^{(5f)}$ = 4.05
and $J^{(5f)}$ = 2.43 for the spin, orbital and total moments. In the
same time, the ground state $|\Omega\rangle$ of the entire impurity
model is a singlet that corresponds to all angular momenta being equal
zero ($S = L = J = 0$). Therefore, the fluctuations in the $f$ shell
are accompanied by canceling antiferromagnetic fluctuations in the
conduction bands, which can be viewed as a manifestation of the Kondo
physics. In addition, our band-structure calculations suggest an
antiferromagnetic instability due to the presence of a Fermi-surface
sheet with a negative second derivative of the Drude plasma energy
(Supplementary Material).

In conclusion, we have performed point-contact Andreev-reflection
measurements in PuCoGa$_5$ single crystals with different values of $T_c$
due to different degrees of disorder. We have shown that PCARS spectra unambiguously indicate that the OP has a $d$-wave symmetry, consistent with indirect
indications from NMR and $\mu$SR measurements. The amplitude of the
gap at $T\rightarrow 0$ is $\Delta = 5.1 \pm 0.3$, as determined from
the fit of different curves, and corresponds to a gap ratio
$2\Delta/k_B T_c$ = 6.4 $\pm$ 0.4. In crystals with Sb impurities and a reduced $T_c$ = 14.5
K, the Andreev signal is greatly reduced, possibly because of a
larger quasiparticle scattering in the superconducting bank arising
from self-induced disorder and impurities. The gap scales with
$T_c$, giving the same gap ratio as in the highest-T$_{c}$ crystals, within the experimental
uncertainty. In both cases, the temperature dependence of the gap
is consistent with the predictions of the
Eliashberg theory for strong-coupling superconductivity if spin
fluctuations provide the mediating bosons.
A characteristic boson energy $\Omega_0=6.5$ meV and
a coupling constant $\lambda=2.37$ allow reproducing the
low-temperature gap values. These results show that PuCoGa$_5$ is an
unconventional, $d$-wave superconductor in the strong-coupling
regime, and indicate spin fluctuations as the probable boson that
mediates superconductivity. Such fluctuations would involve a
time-dependent $5f$ local moment dynamically compensated by a moment
momentarily formed in the surrounding cloud of conduction electrons.

%\pagebreak

\section*{Acknowledgement}
This work has been performed at the Institute of Transuranium Elements
within its "Actinide User Laboratory" program, with
financial support to users provided
by the European Commission.

\section*{Corresponding author}
Correspondence and requests for materials should be addressed to
Roberto Caciuffo, roberto.caciuffo@ec.europa.eu, or to D. Daghero, dario.daghero@polito.it.

\section*{Author contributions}
R.C., D.D, and M. T. designed research; D.D., M.T., and J.-C.G.performed PCARS experiments; R.E., J.-C.G., and E.C. prepared and characterized the
samples; G.A.U. performed Eliashberg calculations, A.B.S., J.K., and
A.I.L. performed {\em ab initio} calculations, D.D., M.T.,
J.-C.G.,E.C., R.C., G.A.U., A.B.S, and J.K. analyzed and interpreted
data; R.C., D.D., M. T., A.B.S., and J.K. wrote the paper.

\section*{Competing financial interests}
The authors declare that they have no competing financial interests.

%\newpage

%\bibliographystyle{naturemag}
%\bibliography{MSWEBpublications}

\newpage

\noindent{\bf Supplementary material for "Strong-coupling
\emph{d}-wave superconductivity in PuCoGa$_5$ probed by point
contact spectroscopy"}

\section{Point contact Andreev reflection spectroscopy}
\label{sect:methods}

Point contact spectroscopy is a simple but very powerful tool for
the investigation of the superconducting order parameter (OP) that,
in the past decades, has been successfully applied, among others, to
cuprates, borocarbides, multiband superconductors and even to the
recently discovered iron-based superconductors. The technique
consists in creating a N-S contact between a normal metal (N) and a
superconductor (S), whose radius $a$ is smaller than both the
electronic mean-free path and the coherence length in S. In these
conditions, an electron travels through the contact ballistically,
i.e. without being diffused, so that if a voltage $V$ is kept at the
junction's end, it enters the superconductor with a maximum excess
energy $eV$. When this energy is smaller than the gap in the
superconductor, $\Delta$, the electron cannot propagate in S as an
electron-like quasiparticle (ELQ), because it finds no available
states. Thus, it forms a Cooper pair and a hole is retro-reflected
in N.

\begin{figure*}
\includegraphics[angle=0,width=17cm]{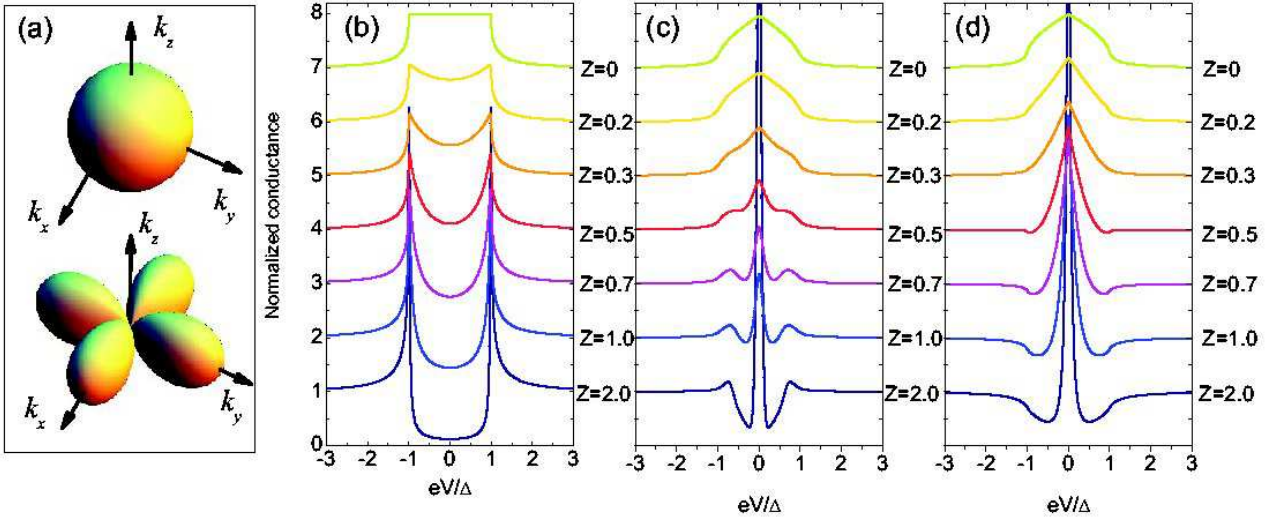}
\caption{(a) Amplitude of the order parameter in spherical
coordinates in the reciprocal space. The cases of a $s$-wave (top)
and of a $d$-wave (bottom) order parameter are shown. (b):
Theoretical normalized conductance curves calculated at $T=0$ for a
$s$-wave gap, and different values of the dimensionless barrier
strength $Z$. (c,d) Normalized conductance curves at T=0 calculated
with the 2D-BTK model for a d-wave gap and assuming $\alpha=\pi/8$
and $\alpha=\pi/4$, respectively; $\alpha$ is the angle between the
direction of the injected current and the $k_{x}$ axis.}
\label{Fig1SM}
\end{figure*}

If the gap is isotropic (Fig. \ref{Fig1SM}a, top) and there is no
potential barrier at the interface, this results in a doubling of
the conductance for $|V|\leq \Delta/e$ (see Fig. \ref{Fig1SM}b, top
curve). If $|eV|>\Delta$, the conductance decreases again towards
the value it would have if the superconductor was in the normal
state. If a barrier is present, other phenomena take place that can
give rise to a normal reflection of the incoming electron, and also
to the transmission of hole-like quasiparticles (HLQ) in S. As a
result, the conductance presents two maxima at approximately $V=\pm
\Delta /e$ and a zero-bias minimum (see Fig. \ref{Fig1SM}b).

If instead the energy gap has a dependence on the direction, i.e.
the gap is anisotropic, electrons injected along different
directions may experience different gap amplitudes. In this case, it
is mandatory to integrate over all the possible directions along
which the electrons can arrive to the interface. In particular, for
a $d_{x^2-y^2}$-wave symmetry of the gap, $\Delta(\theta,
\phi)=\Delta_0 \cos (2\theta)\sin^2(\phi)$, where $\theta$ is the
azimuthal angle in the $k_x,k_y$ plane of the reciprocal space and
$\phi$ the inclination angle (see Fig. \ref{Fig1SM}a, bottom), the
shape of the conductance curves does not depend only on the height
of the potential barrier at the interface, but also on the angle
between the direction of current injection (i.e. the normal to the
interface) and the $k_x$ axis. This angle will be called $\alpha$ in
the following.

The curves shown in Fig. \ref{Fig1SM} were calculated by using the
so called 2D-BTK model \cite{blonder82,kashiwaya96,daghero10} where
the problem is reduced to a two-dimensional one and the Fermi
surface (FS) is supposed to be perfectly cylindrical (with its axis
parallel to the $k_z$ axis). This means that the dependence of the
OP on $\phi$ is disregarded. In this paper, based on the mostly 2D
shape of the largest FS sheets and also for simplicity, we have
always used this model. It can be shown \cite{daghero10} that, with
respect to a more refined 3D model, this approximation generally
gives rise to an overestimation of the parameter $Z$, proportional
to the height of the potential barrier at the interface, which is
not relevant in our analysis.

If $\alpha$ = 0, for any angle of incidence ($\theta$) of the
incoming electron, ELQ and HLQ transmitted in S with angles
$+\theta$ and $-\theta$ feel the same order parameter, in amplitude
and sign. The conductance in this case is doubled only at zero bias
where it shows a characteristic cusp.  If there is no barrier, the
same shape is obtained for any value of $\alpha$: the conductance
curve always looks like the top curve in Fig. \ref{Fig1SM}c. If
instead a barrier is present, for any $\alpha \neq 0$ some values of
$\theta$ exist for which HLQ and ELQ feel order parameters of
opposite sign \cite{daghero10}. This gives rise to constructive
interference between HLQ and ELQ that results in localized
zero-energy states (Andreev bound states). These states manifest
themselves in the conductance giving rise to a peak at zero bias
(ZBCP). When $\alpha=\pi/4$ the current is injected along the nodal
direction, \emph{all} ELQ and HLQ interfere and the ZBCP is maximum.
An example of calculated (normalized) conductance curves assuming
$\alpha=\pi/8$ and $\alpha=\pi/4$ is shown in Figs. \ref{Fig1SM}c
and \ref{Fig1SM}d for increasing values of $Z$.

\section{Solution of Eliashberg equations}
In the imaginary-axis representation, the $d$-wave, one-band
Eliashberg theory \cite{eliashberg60,carbotte90} is formulated by
the following equations for the gap $\Delta_{n}(\phi)
=\Delta(i\omega_{n},\phi)$ and the renormalization functions
$Z_{n}(\phi) = Z\left(i\omega_{n},\phi\right)$:

\begin{eqnarray}
\omega_{n}Z_{n}(\phi)&=&\omega_{n}+\pi T
\sum_{m}\int_{0}^{2\pi}\frac{d\phi'}{2\pi}\lambda_{nm}(\phi,\phi')N^{Z}_{m}(\phi')
\nonumber \\
&+&\gamma_{d} \frac{N^{Z}_{n}}{c^{2}+(N^{Z}_{n})^{2}};
\label{eliashberg1}
\end{eqnarray}

\begin{eqnarray}
&&Z_{n}(\phi)\Delta_{n}(\phi)=\pi T
\sum_{m}\int_{0}^{2\pi}\frac{d\phi'}{2\pi}
\nonumber \\
&&\times[\lambda_{nm}(\phi,\phi')-\mu^{*}(\phi,\phi')\vartheta(\omega_{c}-|\omega_{m}|)]
N^{\Delta}_{m}(\phi'), \label{eliashberg2}
\end{eqnarray}

where $\vartheta$ is the Heaviside function, $\omega_{c}$ is a
cut-off energy for the Coulomb pseudopotential $\mu^{*}$, and
$N^{\Delta}_{m}(\phi)=\Delta_{m}(\phi)
/\sqrt{\omega^{2}_{m}+\Delta_{m}^{2}(\phi)}$, $N^{Z}_{m}(\phi)=
\omega_{m}/\sqrt{\omega^{2}_{m}+\Delta_{m}(\phi)^{2}}$, and
$N^{Z}_{n}$ is an angular average of $N^{Z}_{n}(\phi')$ over the
Fermi surface.

The parameter $\gamma_{d}$ is proportional to the concentration of
defects and $c$ is a parameter related to the electron phase shift
for scattering off an impurity \cite{carbotte94}. The $n^{th}$
Matsubara frequency is defined as $i\omega_{n} = i\pi T(2n-~1)$, and
T is the temperature; $\lambda_{nm}(\phi,\phi') =
\lambda(i\omega_{m}~-~i\omega_{n},~\phi,~\phi')$  is related to the
electron-boson spectral function
$\alpha^{2}(\Omega)F(\Omega,\phi,\phi')$ through the following
equation:

\begin{equation}
\lambda(i\omega_{m}-i\omega_{n},\phi,\phi')=\int_{0}^{+\infty}
\frac{2\Omega\alpha^{2}F(\Omega,\phi,\phi')}{(\omega_{m}-\omega_{n})^{2}+\Omega^{2}}
d\Omega,
\end{equation}

\noindent where $\Omega$ is the boson frequency. We made the usual
lowest-order approximation that both the electron-boson spectral
function and the Coulomb pseudopotential contain separate
\textit{s}- and \textit{d}-wave contributions \cite{rieck90}:

\begin{equation}
\alpha^{2}F(\Omega,\phi,\phi')=\alpha^{2}F_{s}(\Omega)
+2\alpha^{2}F_{d}(\Omega) cos(2\phi) cos(2\phi')
\end{equation}

and

\begin{equation}
\mu^{*}(\phi,\phi')=\mu^{*}_{s} +2\mu^{*}_{d}(\Omega) cos(2\phi)
cos(2\phi')
\end{equation}

Then we assumed for simplicity that
$\alpha^{2}F_{s}(\Omega)=\alpha^{2}F_{d}(\Omega)$, and  that
$\alpha^{2}F_{d}(\Omega)$ has the typical shape of spin fluctuation
spectral functions \cite{millis90}:
$\alpha^{2}F_{d}(\Omega)\propto\frac{\Omega
\Omega_{0}\vartheta(\Omega_{max}-\Omega)}{\Omega^{2}+\Omega_{0}^{2}}$
where $\Omega_{0}$ is the energy peak, $\Omega_{max}$ is a cut off
energy $\Omega_{max}=4 \Omega_{0}$ \cite{monthoux91} and $\omega_c =
3 \Omega_{max}$ .

We then looked for solutions of the equations having pure $d$-wave
symmetry for the gap function
$\Delta(\omega,\phi^{\prime})=\Delta_d(\omega)\sqrt{2}
cos(2\phi^{\prime})$ and pure $s$-wave form for the renormalization
function $Z(\omega,\phi^{\prime})=Z_s(\omega)$; the reason for this
choice is that the only solution of the homogeneous integral
equation for $Z_d(\omega)$ is $Z_d(\omega)=0$, at least for
reasonable values of $\lambda_d$ \cite{musaelian96}. Note that, if a
$d$-wave symmetry for the gap function is assumed, the parameter
$\mu_s^{*}$ does not enter into the two relevant Eliashberg
equations. Therefore, although it is certainly larger than
$\mu_d^{*}$, it does not influence the solution. The free parameters
are then: $\gamma_{d}$, $c$, $\Omega_{0}$, $\mu_d^{*}$ and
$\lambda=\int_{0}^{+\infty}
\frac{2\Omega\alpha^{2}F_{d}(\Omega)}{\Omega} d\Omega$.

The parameters $c$ and $\gamma_{d}$ were estimated by reanalyzing
the normalized local spin susceptibility from the NMR experiment by
Curro et al. \cite{curro05}; in particular, it was necessary to
assume $\gamma_{d} = 0.25 ~ {\rm meV}$ and $c=0$ (unitary limit) to
fit the shape of the curve \cite{jutier08}. The unitary limit is the
only possible choice to reproduce the dependence of the critical
temperature on disorder \cite{jutier08}. As for $\mu_d^*$, it
generally ranges between 0 and 0.2 \cite{carbotte90}. Finally, let
us mention here that a range for $\Omega_0$ was recently provided by
NMR \cite{baek10} measurements: $4 \leq \Omega_{0}\leq 8$ meV.

\begin{figure}[htbp]
\includegraphics[angle=0,width=15cm]{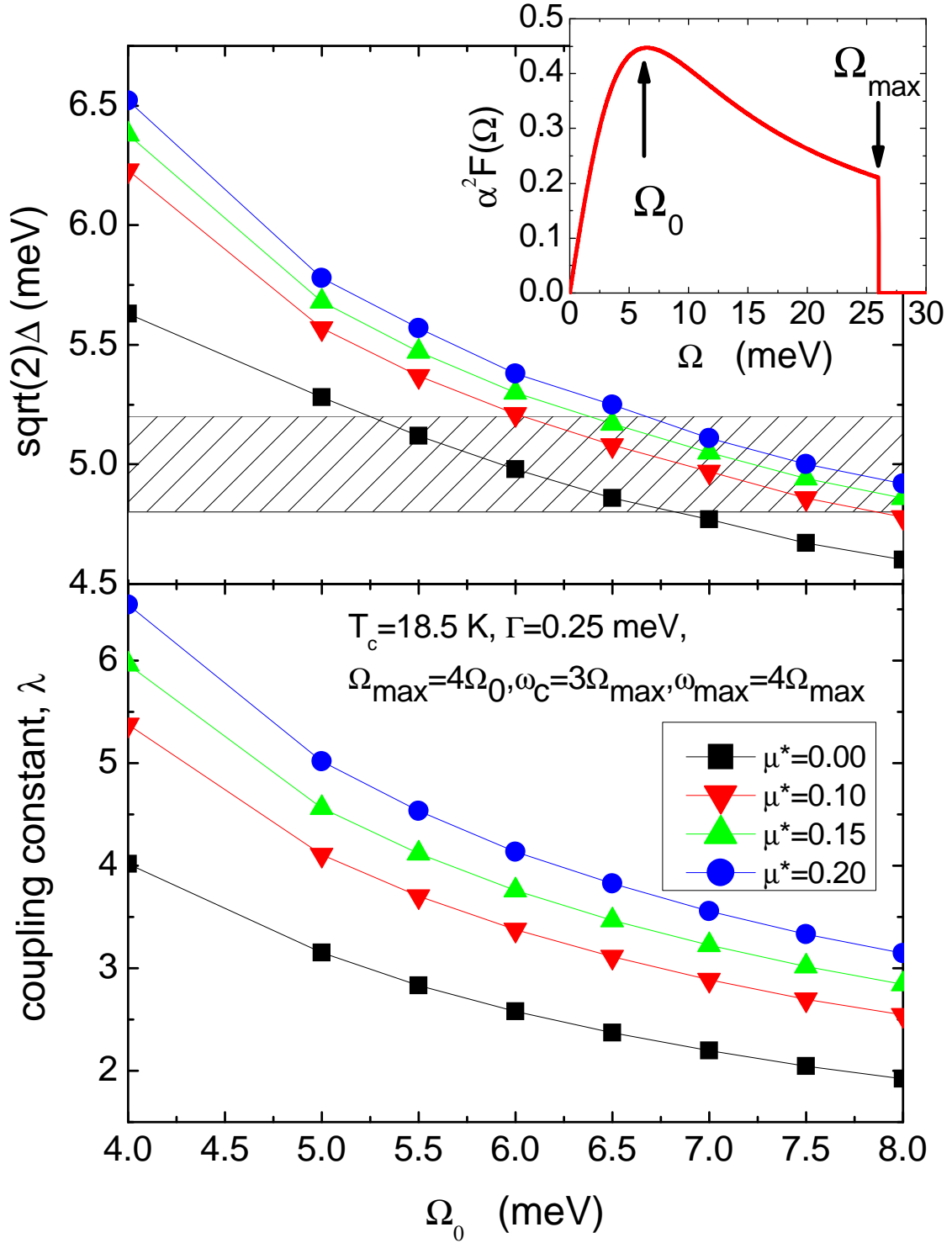}
\caption{(a) The effect of $\Omega_{0}$ on $\Delta$  for different
values of the Coulomb pseudopotential $\mu_d^*$. The dashed region
indicates the range of gap values extracted from the point contact
Andreev reflection measurements. The inset shows the shape of the
electron-spin-fluctuation spectrum. (b) The effect of $\Omega_0$ on
the coupling constant $\lambda$ for different values of the Coulomb
pseudopotential.}\label{Fig2SM}
\end{figure}

The value of $\lambda$ was chosen so as to obtain the experimental
critical temperature ($T_{c}=18.5$ K) by solution of the
imaginary-axis Eliashberg equations, and then the experimental
low-temperature gap $\Delta$ by solution of the real-axis Eliashberg
equations \cite{ummarino10} for different values of $\Omega_{0}$ and
$\mu_d^{*}$. Fig. \ref{Fig2SM}a shows the values of the gap as a
function of $\Omega_0$ for different values of $\mu_d^*$. The
corresponding values of $\lambda$ are shown in Fig. \ref{Fig2SM}b.
From the results of point-contact Andreev reflection measurement we
know that, in PuCoGa$_{5}$  single crystals with $T_{c}$ =18.5 K the
low-temperature gap lies in the range 4.8-5.2 meV. As we can see
from Fig. \ref{Fig2SM}a, this corresponds to an interval of possible
values of $\Omega_0$ that goes from 5.3 to about 8.9. From the
intersection of this range with that given by NMR measurement
\cite{baek10} one obtains $5.3 \leq \Omega_{0}\leq 8$ meV. Looking
at Fig. \ref{Fig2SM}b, it is clear that this corresponds to $2.2
\leq \lambda \leq 3.7$.

To reproduce the temperature dependence of the gap, we chose for
simplicity  $\mu_d^{*}$ = 0 and adjusted the values of $\Omega_{0}$
and $\lambda$. We found that the experimental low-temperature gap
and the $T_c$ can be obtained with $\Omega_{0}$ = 6.5  meV and
$\lambda$ = 2.37.

In order to obtain $T_{c}$ =14.5 K it is necessary to increase the
scattering parameter to $\gamma_{d}$ =1.6 meV. In this way, we were
able to reproduce the temperature dependence of the gap in the
low-$T_c$ samples with the same values of $\Omega_0$ and $\lambda$
mentioned above (see Fig. \ref{oldcrystals}c). This means that all
the difference between the two sets of data can be ascribed to the
greater amount of impurity scattering in the lower-$T_c$ crystals.

\section{Electronic Structure Calculations}
Following Ref.~\cite{LK99} we consider the multi-band Hubbard
Hamiltonian $H = H^0 + H^{\rm int} $, where
\begin{eqnarray}
\label{eq:1ph} H^0 = \sum_{i,j} \sum_{\gamma_1, \gamma_2} H^0_{i
\gamma_1, j \gamma_2}
                 c^{\dagger}_{i \gamma_1} c_{j \gamma_2}
    = \sum_{\bf k} \sum_{\gamma_1, \gamma_2} H^0_{\gamma_1, \gamma_2} ({\bf k})
        c^{\dagger}_{\gamma_1}({\bf k}) c_{\gamma_2}({\bf k})
\end{eqnarray}
is the one-particle Hamiltonian found from \textit{ab initio}
electronic structure calculations of a periodic crystal. The indices
$i,j$ label lattice sites, $\gamma = (l m \sigma)$  mark
spinorbitals $\{ \phi_{\gamma} \}$, and ${\bf k}$ is a vector from
the first Brillouin zone. We assume that the electron-electron
correlations between $s$, $p$, and $d$ electrons are well
approximated within DFT, and consider the correlations between the
$f$ electrons introducing the interaction Hamiltonian
\begin{eqnarray}
\label{eq:hint} H^{\rm int} = \frac{1}{2} \sum_{i}
\sum_{m_1,m_2,m_3,m_4}^{\sigma,\sigma'}
  \langle m_1,m_2|V_{i}^{ee}|m_3,m_4 \rangle c_{i m_1 \sigma}^{\dagger} c_{i m_2 \sigma'}^{\dagger}
  c_{i m_4 \sigma'} c_{i m_3 \sigma}  \, .
\end{eqnarray}
The operator $V^{ee}$ represents an effective on-site Coulomb
interaction \cite{LK99} expressed in terms of the Slater integrals
$F_k$ and the spherical harmonics ${|lm \rangle}$.

In what follows, the effects of the interaction Hamiltonian $H^{\rm
int}$ on the electronic structure are modeled by a ${\bf
k}$-independent one-particle selfenergy $\Sigma(z)$, which is
constructed with the aid of an auxiliary impurity model describing
the complete seven-orbital $f$ shell including the full spherically
symmetric Coulomb interaction, spin-orbit coupling (SOC), and
crystal field (CF). The Hamiltonian of this multi-orbital impurity
model can be written as \cite{Hewson}
\begin{align}
\label{eq:hamilt} H_{\rm imp}  = & \sum_{\substack {k m m' \\ \sigma
\sigma'}}
 [\epsilon^{k}]_{m m'}^{\sigma \; \; \sigma'} b^{\dagger}_{km\sigma}b_{km'\sigma'}
 +\sum_{m\sigma} \epsilon_f f^{\dagger}_{m \sigma}f_{m \sigma}
\nonumber \\
& + \sum_{mm'\sigma\sigma'} \bigl[\xi {\bf l}\cdot{\bf s}
  + \Delta_{\rm CF}\bigr]_{m m'}^{\sigma \; \; \sigma'}
  f_{m \sigma}^{\dagger}f_{m' \sigma'}
\nonumber \\
& +  \sum_{\substack {k m m' \\ \sigma \sigma'}}   \Bigl( [V^{k}]_{m
m'}^{\sigma \; \; \sigma'}
 f^{\dagger}_{m\sigma} b_{km' \sigma'} + \text{h.c.}
  \Bigr)
\\
& + \frac{1}{2} \sum_{\substack {m m' m''\\  m''' \sigma \sigma'}}
  U_{m m' m'' m'''} f^{\dagger}_{m\sigma} f^{\dagger}_{m' \sigma'}
  f_{m'''\sigma'} f_{m'' \sigma},
\nonumber
\end{align}
where $f^{\dagger}_{m \sigma}$ creates an electron in the $f$ shell
and $b^{\dagger}_{m\sigma}$ creates an electron in the ``bath''
which consists of those host-band states that hybridize with the
impurity $f$ shell. The impurity-level position $\epsilon_f$ and the
bath energies $\epsilon^{k}$ are measured from the chemical
potential $\mu$. The parameters $\xi$ and $\Delta_{\rm CF}$ specify
the strength of the SOC and the size of the CF at the impurity. The
parameter matrices  $V^{k}$ describe the hybridization between the
$f$ states and the bath orbitals at the energies $\epsilon^{k}$.

The band Lanczos method~\cite{ruhe1979,meyer1989} is employed to
find the lowest-lying eigenstates of the many-body Hamiltonian
$H_{\rm imp}$ and to calculate the one-particle Green's function in
the subspace of the $f$ orbitals $[G_{\rm imp}(z)]_{m m'}^{\sigma \;
\; \sigma'}$ at low temperature ($k_{\rm B}T=1/500$ eV). The sought
for selfenergy $[\Sigma (z)]_{m m'}^{\sigma \; \; \sigma'}$ is then
straightforwardly obtained from the inverse of the Green's function
matrix $[G_{\rm imp}(z)]_{m m'}^{\sigma \; \; \sigma'}$. Note that
for the Lanczos method to be applicable, the continuum of the bath
states is discretized. In particular, the calculations presented
here utilize only a single value for the index $k$.

Once the selfenergy is known, the local Green's function $G(z)$ for
the electrons in the solid is calculated as
\begin{equation}
%\hspace*{-0.5cm}
[G(z)]^{-1}_{\gamma_1 \gamma_2}  = [{G}_{\rm LDA}(z)]^{-1}_{\gamma_1
\gamma_2}
 - \Delta \epsilon\, \delta_{\gamma_1 \gamma_2} -
[\Sigma(z)]_{\gamma_1 \gamma_2},\label{eq:local_gf}
\end{equation}
where $\Delta \epsilon$ accounts for the difference between the
impurity and the lattice chemical potentials,
% and is chosen to fix the number of $f$~electrons $n_f$,
and $G_{\rm LDA}(z)$ is the LDA Green's function,
\begin{equation}
[G_{\rm LDA}(z)]_{\gamma_1 \gamma_2} = \frac{1}{V_{\rm BZ}}
\int_{\rm BZ}{\rm d}^3 k \,\bigl[z+\mu-H_{\rm LDA}({\bf
k})\bigr]^{-1}_{\gamma_1 \gamma_2}\,. \label{eq:gf_lda}
\end{equation}
For the charge-density self-consistency, we employ the so-called
local density matrix approximation (LDMA)~\cite{shick09}. In the
LDMA, the occupation matrix
\begin{equation}
n_{\gamma_1 \gamma_2} = -\frac1{\pi}\,\mathop{\rm Im}
\int_{-\infty}^{E_{\rm{F}}} {\rm d} z \, [G(z)]_{\gamma_1 \gamma_2}
\label{eq:occmtx}
\end{equation}
is self-consistently evaluated with the aid of the local Green's
function $G(z)$ from Eq.~(\ref{eq:local_gf}). The matrix
$n_{\gamma_1 \gamma_2}$ is then used to construct an effective
LDA+$U$ potential ${V}_{U}$ which is inserted into Kohn--Sham-like
equations,
\begin{gather}
%\Big( -\nabla^{2} + V(\mathbf{r})_{\rm LDA} &+& V_{U} - V_{DC} + \xi
[ -\nabla^{2} + V_{\rm LDA}(\mathbf{r}) + V_{U} + \xi ({\bf l} \cdot
{\bf s}) ]  \Phi_{\bf k}^b({\bf r}) = \epsilon_{\bf k}^b \Phi_{\bf
k}^b({\bf r}) .
%\;, \nonumber \\
%\rho(\mathbf{r}) = \sum_i^{occ} \Phi^{\dagger}_i({\bf r})\Phi_i({\bf
%r}) \, ,
\label{eq:kohn_sham}
\end{gather}
These equations are iteratively solved until self-consistency over
the charge density is reached. In each iteration, a new Green's
function ${G}_{\mathrm{LDA}}(z)$ and a new value of the $f$-shell
occupation are obtained from the solution of
Eq.~(\ref{eq:kohn_sham}). Subsequently, a new selfenergy $\Sigma(z)$
corresponding to the updated $f$-shell occupation is constructed.
Finally, the next iteration is started by inserting the new
${G}_{\mathrm{LDA}}(z)$ and $\Sigma(z)$ into
Eq.~(\ref{eq:local_gf}). After the iterations are completed, the
imaginary part of the local Green's function $G(z)$ provides a means
to estimate the valence-band photoemission (PE) spectra.

In the calculations we used  the in-house
implementation~\cite{shick99,shick01} of the full-potential
linearized augmented plane wave (FP-LAPW) method~\cite{wimmer}. This
FP-LAPW version includes all relativistic effects:
scalar-relativistic and spin-orbit coupling. The calculations were
carried out assuming a paramagnetic state with the LDA
optimized~\cite{opahle03}  crystal structure parameters of
PuCoGa$_5$. The Slater integrals were chosen as $F_0=4.0$~eV, and
$F_2=7.76$ eV, $F_4=5.05$ eV, and $F_6=$ 3.07 eV~\cite{Moore}. They
corresponds to commonly accepted values for Coulomb~$U=4.0$ eV and
exchange~$J = 0.64$ eV. The SOC parameter was determined from LDA
calculations as $\xi=0.28$ meV. The crystal-field effects were found
to be negligible and we set $\Delta_{\rm CF}=0$.

\begin{figure}[htbp]
\includegraphics[angle=0,width=12cm]{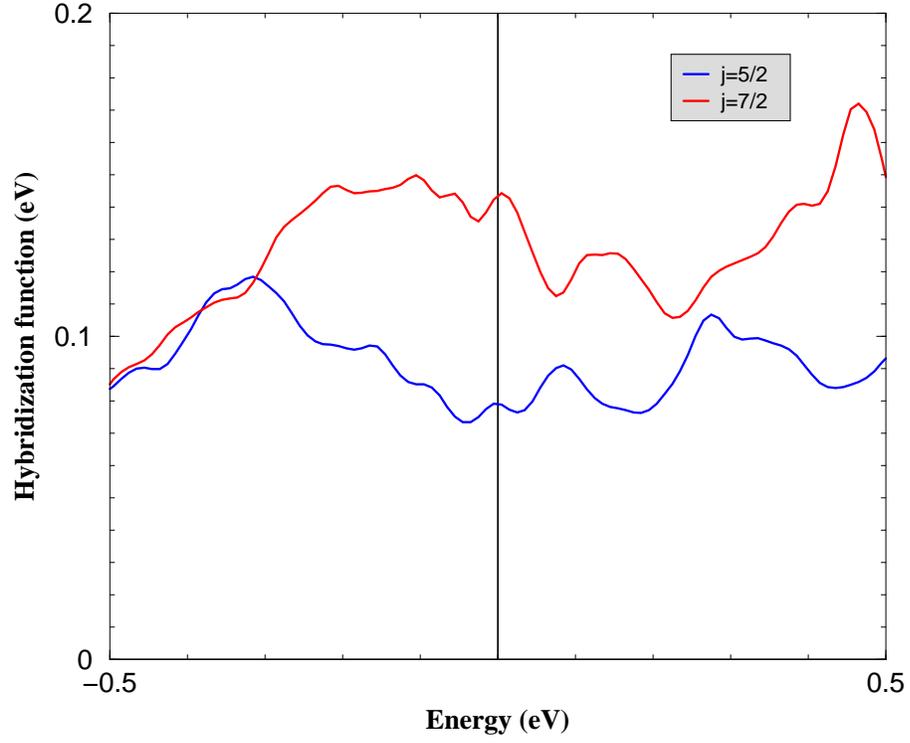}
\caption{(Color online)  LDA hybridization function ${\Delta} = {1
\over {\pi N_f}} Im Tr [G^{-1}(\epsilon + i \delta)]$ for $j=5/2$
and $7/2$ and $\delta$ = 31.4 meV.}  \label{hybridization}
\end{figure}

The first and fourth terms in the impurity model,
Eq.~(\ref{eq:hamilt}), are to a good approximation diagonal in the
$\{j,j_z\}$ representation, so that we only need to specify one bath
state (six orbitals) with $\epsilon_{k=1}^{j=5/2}$ and
$V_{k=1}^{j=5/2}$, and another bath state (eight orbitals) with
$\epsilon_{k=1}^{j=7/2}$ and $V_{k=1}^{j=7/2}$. In order to
determine these bath parameters, we assume that LDA represents the
non-interacting model for PuCoGa$_5$, that is, we associate the LDA
solution with the Hamiltonian of Eq.~(\ref{eq:hamilt}) where we set
the Coulomb interaction matrix $U_{mm'm''m'''}=0$. The numerical
values of the bath parameters are then found as follows: We
calculate the LDA hybridization function ${\Delta} = {\frac1{\pi
N_f}} \mathop{\rm Im}\mathop{\rm Tr} [G^{-1}_{\rm LDA}(\epsilon + i
\delta)]$ for $j=5/2$ ($N_f=6$) and for $j=7/2$ ($N_f=8$), which is
shown in Fig.~\ref{hybridization}. Assuming that the most important
hybridization is the one occurring in the vicinity of $E_F$, we
obtain $V_{k=1}^{5/2,7/2}$ from the relation $\sum_{k}
{|V_{k}^{j}|}^2 \delta(\epsilon_{k}^{j} - \epsilon) =
\Delta^{j}(\epsilon)$~\cite{Gunnarson} averaged over the energy
interval $\epsilon\in (E_F-0.5\,{\rm eV},E_F+0.5\,{\rm eV})$. The
bath-state energies $\epsilon_{k=1}^{5/2,7/2}$ are adjusted to
approximately reproduce the LDA $5f$-states occupations $n_f^{5/2}$
and $n_f^{7/2}$.

\begin{table}[htbp]
%\vspace*{-0.25cm}
\caption{$5f$-states occupations $n_f$, $n_f^{5/2}$ and $n_f^{7/2}$,
$\epsilon_1^{5/2,7/2}$  (eV), $V_{1}^{5/2,7/2}$ (eV) for
PuCoGa$_5$.} \label{occup}
\begin{ruledtabular}
\begin{tabular}{lccccccc}
 $n_f$  & $n_f^{5/2}$ &$n_f^{7/2}$&$\epsilon_1^{5/2}$&$V_{1}^{5/2}$&$\epsilon_1^{7/2}$&$V_{1}^{7/2}$   \\
\hline
 5.14 & 4.38 & 0.76 & 0.25 & 0.29   & -0.07 &0.34\\
\end{tabular}
\end{ruledtabular}
\end{table}

% After the parameters of the discrete impurity model are set, the band
% Lanczos method~\cite{ruhe1979,meyer1989} is utilized to determine the
% lowest lying eigenstates of the many-body Hamiltonian and to calculate
% $\Sigma(z)$.  The inverse temperature $\beta=500$ eV$^{-1}$ was used
% in these calculations. The local density matrix Eq.(~\ref{eq:occmtx})
% is evaluated, and used to construct the $V_{U}$ potential.

In what follows, we have adopted the fully-localized (or
atomic-like) limit (FLL)  for the double-counting term $V_{dc} = U
(n_f-1/2) -
%J(n^{\sigma}-\frac{1}{2})
J(n_f-1)/2$ entering the definition of the LDA+$U$ potential $V_U$.
% The charge density self-consistency is performed by solving
% iteratively the Eq.(~\ref{eq:kohn_sham}) in FP-LAPW basis.
In the FP-LAPW calculations we set the radii of the atomic spheres
to 3.1~a.u.~(Pu), 2.3~a.u.~(Co) and 2.3~a.u.~(Ga). The parameter $R
\times K_{\text{max}}=10.2$ determined the basis set size, and the
Brillouin zone (BZ) sampling was performed with 1152 $k$~points. The
self-consistent procedure defined by
Eqs.~\eqref{eq:local_gf}--\eqref{eq:kohn_sham} is repeated until the
convergence of  the $f$-manifold occupation $n_f$ is better than
0.01.

The resulting $f$-orbital DOS (or, the one-particle spectral
function) is shown in Fig.~4a of the main text. The DOS below $E_F$
is in reasonably good agreement with Ref.~\cite{pezzoli11}. Above
$E_F$,  the DOS obtained here differs somewhat from the DOS reported
in Ref.~\cite{pezzoli11} for a reason which is unclear at the
moment. We estimate the electronic specific heat coefficient using
the formula $\gamma= \frac{\pi^2}{3}k_B^2 \mathop{\rm
Tr}[N(E_F)(1-\frac{d
  \Sigma(\omega)}{d\omega})|_{\omega=0}]$ and we get
$\approx$ 40 mJ K$^{-2}$ mol$^{-1}$ which is smaller than the
experimentally estimated value of 80--100 mJ K$^{-2}$ mol$^{-1}$.
Note that a possible enhancement of $\gamma$ due to the
electron-phonon interaction is not taken into account.

% The valence histogram  calculated by projecting the ED-ground state
% $|\Omega\rangle$ onto Pu atomic eigenvectors is shown in Fig. 4b
% of the main text. Here, the $n_f$-weight (probability) $P_m$ defines
% the f-valence $<n_f> = \sum_{m} P_m n_m$, where $n_m$ is an integer
% number of f-electrons in the m$^{th}$-state. The highest
% probability is $P_5$=0.63, the next is $P_6=0.33$. In addition, there
% are non-zero probabilities for $P_4$=0.03 and $P_7$=0.01. This
% suggests a mixed-valence nature of the $f$-shell for the Pu atoms, which
% consists of a mixture of $f^5$ and $f^6$ multiplets similar to
% $\delta-Pu$~\cite{shim07}. This mixed (or intermediate) valence is
% consistent with a large value of the $\gamma$-coefficient which is
% typically observed in mixed-valence rare-earth based materials.

Now we discuss the magnetic properties. The ground state of the
cluster formed by the $f$~shell and the bath, Eq.~(\ref{eq:hamilt}),
is a singlet ($S=L=J=0$) that includes $\langle n_f \rangle=5.30$ in
the $f$~shell, and $\langle n_{bath} \rangle=8.70$ in the bath
states. The spin $S=2.18$, orbital $L=4.05$ and total $J=2.43$
moments are calculated for the $f$ shell from the expectation values
$\langle \hat X^2 \rangle=X(X+1)$, $X=S,L,J$. The individual
components of the moments vanish, $\langle \hat S_z \rangle =
\langle \hat L_z \rangle=0$, unless the symmetry is broken by an
external magnetic field. We carry out calculations with a small
external magnetic field added to Eq.~(\ref{eq:hamilt}) in order to
estimate the local $f$-shell susceptibility. These calculations
yield $\approx 0.3 \times 10^{-3}$ $\mu_B/$T which is somewhat
smaller than the experimental value $0.74 \times 10^{-3}$
$\mu_B/$T~\cite{hiess08}.

Qualitatively, the non-magnetic character of the composite electron
cluster is linked to the mixed-valence nature of the Pu atom. The
fluctuating local moment is compensated by corresponding dynamical
fluctuations of the bath moment. The magnetic susceptibility $\chi
\sim \frac{1}{(T + T_{fc})}$ is anticipated, similar to a
Kondo-singlet state, which remains constant for $T<<T_{fc}$, as
observed experimentally \cite{hiess08}. Note that neither the
admixture of atomic-like $f^5$ and $f^6$ multiplets nor the
Kondo-like screening of the  $f^5$ state localized moment
\cite{Matsumoto11} by the bath alone are sufficient  to form a
singlet ground state of the composite cluster formed by the f-shell
and the bath. When the hybridization is set to zero in
Eq.~(\ref{eq:hamilt}) (LDA+Hubbard-I approximation~\cite{shick11})
the resulting mixed-valence solution with $\langle n_f \rangle$ =
5.3 is not a singlet and the f-shell stays magnetic. On the other
hand, when we force $f^5$ solution in Eq.~(\ref{eq:hamilt}) and keep
hybridization unchanged, the ground state of the composite electron
cluster is not in a singlet either. In this sense, the admixture of
the non-magnetic $f^6$ to the magnetic $f^5$ state enhances the
dynamical screening of the local moment by the bath.

\begin{figure}[htbp]
\includegraphics[angle=0,width=0.925\columnwidth,clip]{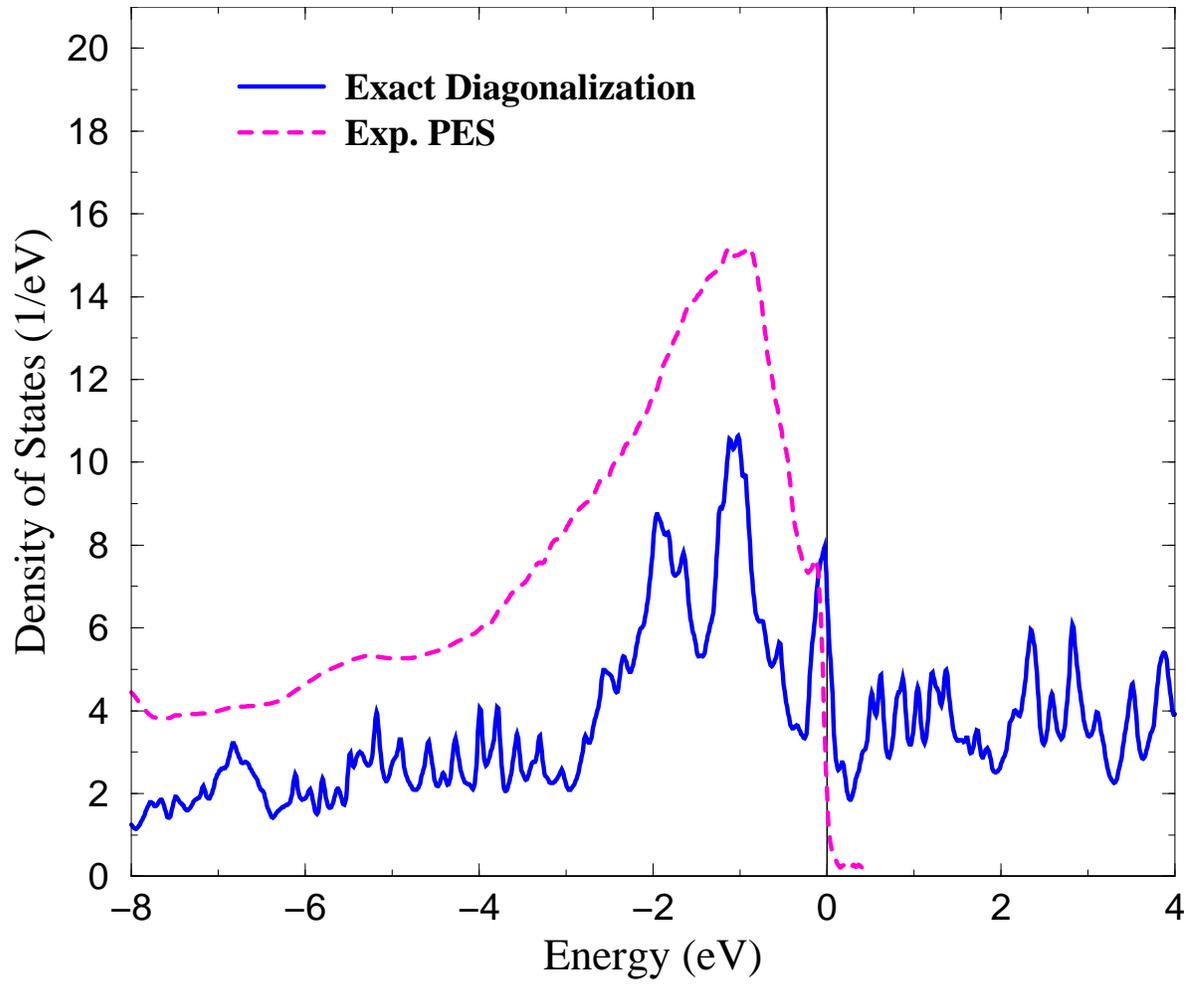}
 \caption{(Color online) Total DOS  of
    PuCoGa$_5$ provided by LDA+ED with $U=4.0$~eV. The experimental PE
    spectrum (arb. units) is taken from Ref.~\cite{eloirdi09}. \label{fig:pes}}
\end{figure}

Comparison of theoretical and experimental PE spectra represents an
important criterion of soundness of the electronic structure
calculations. The use of the single-particle LDA DOS to compare with
the PE spectrum is justified only in case of weak electron
correlations among the $5f$ states. For the $5f$ states at the
borderline between the localized and itinerant behavior neither the
LDA nor the static-mean-field LDA+U theories are sufficient to
accurately describe the PE since these approximations entirely miss
the experimentally observed atomic-like multiplets. In
Fig.~\ref{fig:pes} we show the total DOS for PuCoGa$_5$ as provided
by our LDA+ED calculations. Overall, the agreement with the
experimental PE data \cite{eloirdi09} is improved over LDA+Hubbard-I
calculations reported earlier~\cite{shick11}.

\begin{center}
\begin{figure}[htbp]
%\vspace*{-1cm}
\hspace*{0cm}
\includegraphics[angle=0,width=17cm]{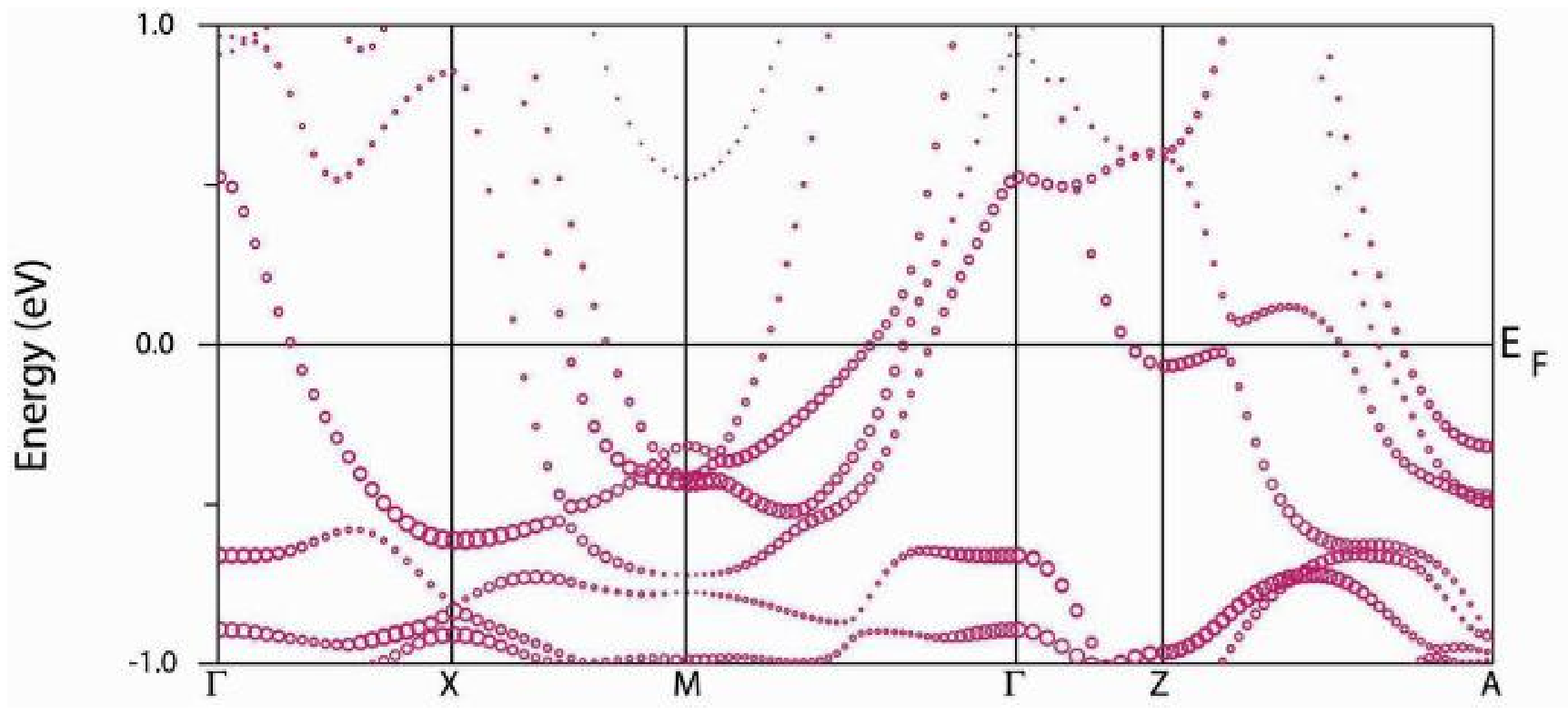}
\hspace*{0cm}
\includegraphics[angle=0,width=17cm]{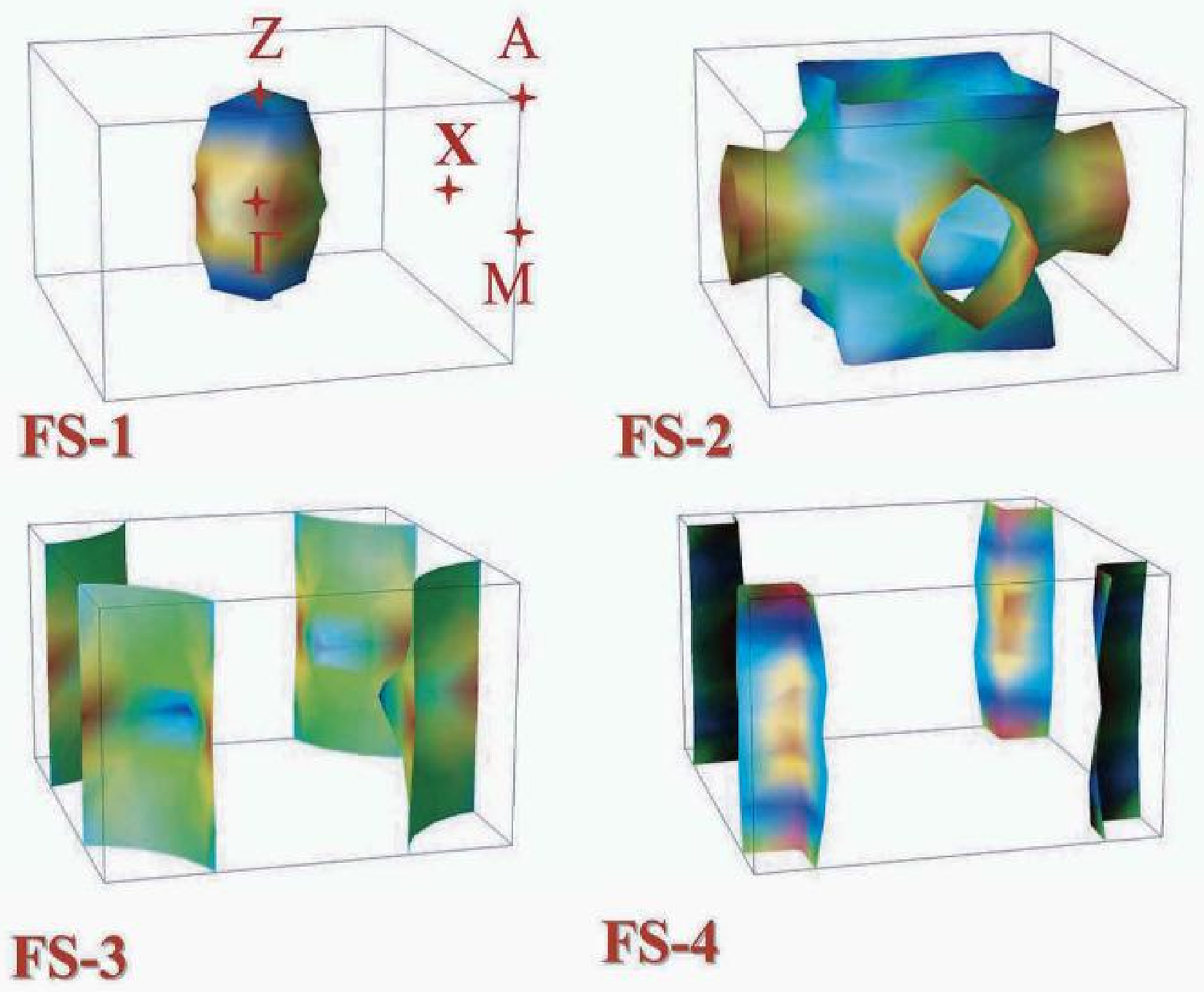}
\caption{(Color
online) (Top) the band structure; (bottom) Fermi surface of
PuCoGa$_5$ for $U=4$~eV obtained from LDA+ED calculations.
\label{fig:ldma_bands}}
\end{figure}
\end{center}

Now we discuss the static band-structure resulting from the
solutions of Eq.~(\ref{eq:kohn_sham}). In fact, as it can be seen
from Fig. \ref{fig:ldma_bands} (upper panel), near $E_F$ it is very
similar to the static band structure obtained from LDA+HIA
calculations~\cite{shick11}. The corresponding quasi-particle Fermi
surfaces (FS) from Ref.~\cite{shick11} are shown in the lower panel
of Fig.~\ref{fig:ldma_bands}. There are four sheets (I--IV)
composing the FS: sheets I and II are fairly three-dimensional and
sheets III and IV are two-dimensional. Close similarities are
revealed between these sheets and the Fermi surfaces from the
previous LDA \cite{opahle03} and AMF-LDA+U calculations. As it was
already noticed in Ref.~\cite{shick11}, FS calculated with LDMA and
LDA have very similar geometry.

Now we calculate the Drude plasma energy,
$\Omega_{p}=\sqrt{\frac{1}{3} \Omega_x^2 + \frac{1}{3} \Omega_y^2 +
\frac{1}{3} \Omega_z^2}$
%\end{equation}
where the individual directional components $\Omega_{i}$ are given
by
\begin{equation}
\Omega_{i}^2 = \frac{e^2}{2 \pi^2} \int_{\rm BZ} {\rm d}^3{k}\,
v_i^2 \delta \bigl(E(\mathbf{k})-E_F\bigr)\,,
\end{equation}
which leads to numerical value of $\Omega_p=3.8$~eV, as in previous
LDA+HIA calculations~\cite{shick11}.  Eventually, it leads to the
same estimate for the electron-phonon coupling $\lambda_{\text{tr}}
\approx 2.5$ from the Bloch--Gr\"uneisen transport theory.

In addition, we can estimate the type of the magnetic instability,
ferro- or antiferromagnetic, for the band structure in
Fig.~\ref{fig:ldma_bands}. The paramagnetic susceptibility
$\chi_0(q,\omega)$ in the limit of small $q$ and $\omega$ can be
expanded~\cite{larson2004} as $\chi_0(q,\omega) = N(E_F)- aq^2 + \i
b {\omega \over q}$. In a simplified form~\cite{jeong2006}, the
coefficient $a$ is proportional to a second derivative ${d^2
\Omega^2_p(E) \over d^2 E} |_{E=E_F}$ of the Drude plasma energy.
The sign of this derivative indicates the type of magnetic
instability. A positive sign corresponds to $q=0$ ferromagnetic
fluctuations, whereas a negative sign to $q \neq 0$
antiferromagnetic (or spin-spiral) fluctuations. In Table II we show
the second derivative of $\Omega_P(E)$ for all four FS bands
calculated from cubic polynomial fits. Our estimate points to FS-II
as a suitable candidate for $q \neq 0$ antiferromagnetic
instability. The electrons at FS-II, at least in principle, can be
paired by anti-ferromagnetic fluctuations.

It is also to note that the shape of FS shown in the lower panel of
Fig.~\ref{fig:ldma_bands} suggests a possibility of nesting between
the hole-like FS-1 around the $\Gamma=(0,0,0)$ point and
electron-like FS-3 around the $M=(\pi,\pi,0)$ point. This large $q
\sim (\pi,\pi,0)$ nesting-like feature can lead to a peak in the
spin susceptibility, and promote the anti-ferromagnetic fluctuations
to develop. This suggest similarities between the coupling
mechanisms for the high-$T_c$ superconductivity in PuCoGa$_5$ and
Fe-based superconductors \cite{mazin08}.

\begin{table}[htbp]
\caption{The total DOS($E_F$) in eV${}^{-1}$, Drude plasma energy
(eV), and its second derivative over energy at $E_F$}
\begin{tabular}{cccccc}
\hline
&& FS-I & FS-II & FS-III & FS-IV \\
\hline
\multicolumn{2}{c} {DOS($E_F$) (1/eV)}& 0.7 & 1.3 & 0.8 & 0.5 \\
 \multicolumn{2}{c} {$\Omega_p(E_F)$ (eV)}& 1.1 & 2.7 & 1.9  & 1.5\\
 \multicolumn{2}{c} {${d^2 \Omega^2_p(E_F) \over d^2 E_F}$}& 49.71 & -221.14 & 72.0 & 193.7 \\
\hline
\end{tabular}
\end{table}

\section{Characterization of the investigated samples}
\begin{figure}[htbp]
\includegraphics[angle=0,width=12cm]{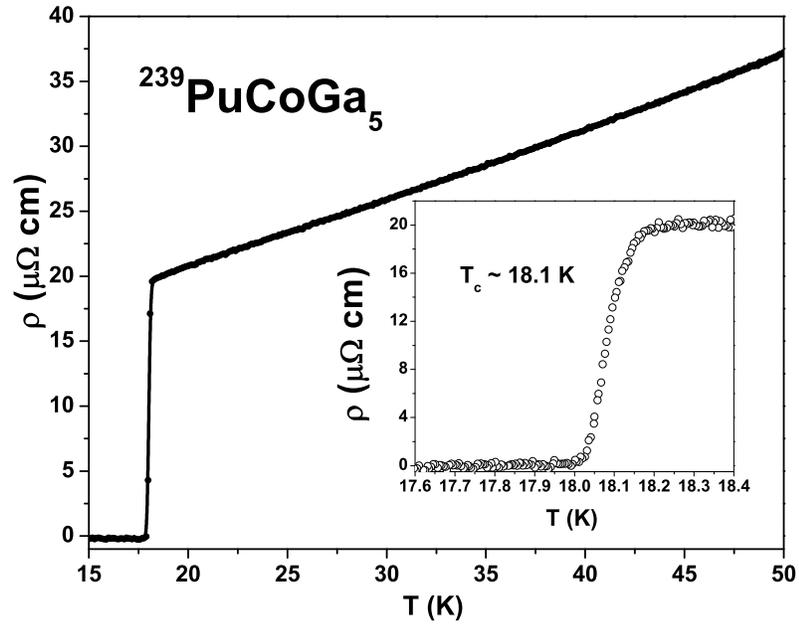}
\caption{Temperature dependence of the electrical resistivity of a
$^{239}$PuCoGa$_{5}$ single crystal after thermal annealing of the
self-radiation damage.} \label{Pu239}
\end{figure}

The samples used for these experiments have been characterized by
magnetic susceptibility and electrical resistivity measurements,
using commercial Quantum Design platforms (MPMS 7-T SQUID, and
PPMS-9T). $^{239}$PuCoGa$_{5}$ single crystals have been submitted
to a thermal treatment to anneal the self-radiation damage.
Fig.~\ref{Pu239} shows the temperature dependence of the electrical
resistivity measured for one of these crystals immediately after a
thermal treatment. All the samples exhibited a critical temperature
very close to the optimal value T$_{c}$ = 18.5 K. Point Contact
Andreev Reflection Spectroscopy (PCARS) measurements were started
within one day from the thermal treatment. PCARS measurements have
also been performed on
 $^{242}$PuCoGa$_{5}$ crystals featuring a lower $T_c$ =14.5 K, due to the presence of Sb impurities (less than 1 \%). The results obtained for these samples are presented in the following section.

\section{PCARS measurements on samples with $T_c$ =14.5 K}
\begin{figure}[htbp]
\includegraphics[angle=0,width=18cm]{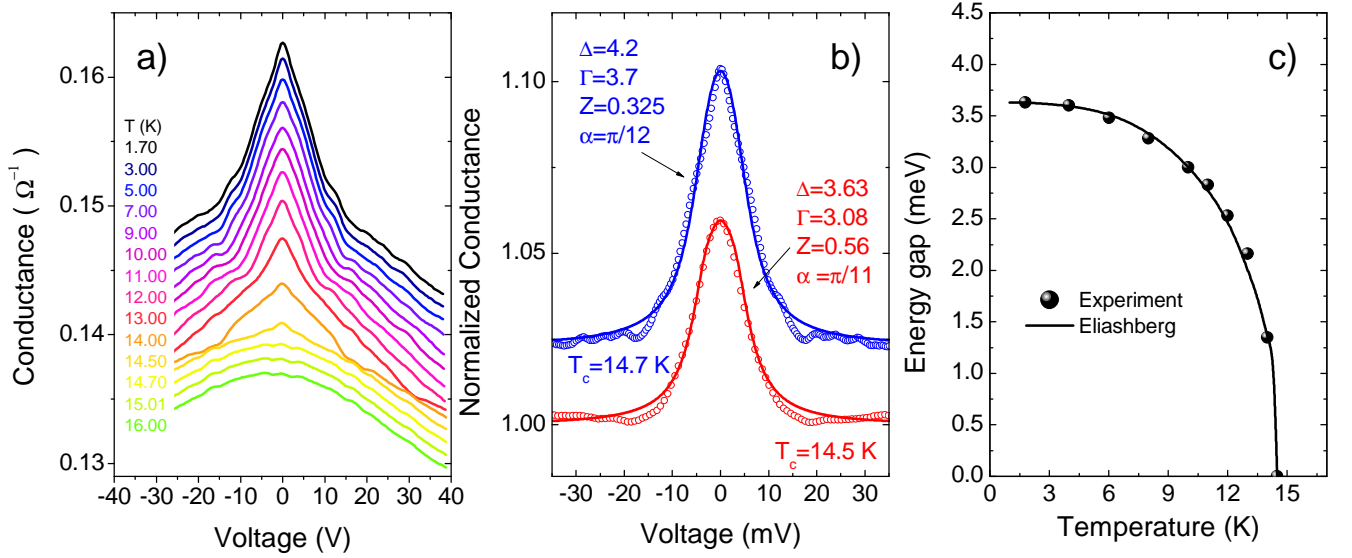}
\caption{(a) Temperature dependence of the conductance curves of a
contact with $R_N=6.9 \Omega$ on a $^{242}$PuCoGa$_{5}$ crystal with
reduced $T_c$ =14.5 K. The curves are vertically offset for clarity.
(b) Two examples of low-temperature normalized conductance curves in
two contacts on the same crystals (symbols). Lines represent the
best-fit of the spectra within the d-wave 2D-BTK model. The
normal-state resistance is $R_N=6.9$ $\Omega$ (top) and $R_N=18.8$
$\Omega$ (bottom). (c) Temperature dependence of the gap extracted
from the fit of the conductance curves of the contact with
$R_N=18.8$ $\Omega$ (symbols). The solid line is the corresponding
$\Delta(T)$ function calculated within the Eliashberg theory by
assuming $\Omega_0=6.5$ meV and a coupling constant $\lambda=2.37$.}
\label{oldcrystals}
\end{figure}

The point contacts were made on the side surface of the crystals so
as to inject the current mainly along the $ab$ planes. Unlike in the
purest samples, here we used a small spot of Ag conducting paste
between the Au wire and the sample to act as the N electrode and
also to mechanically stabilize the contacts. Fig. \ref{oldcrystals}a
shows a series of conductance curves measured as a function of
temperature in a contact with $R_N= 6.9 \Omega$. The curves are
vertically offset for clarity; as in the cases discussed for the
sample with T$_{c}$ = 18.5 K, there is no shift of the tails on
increasing temperature.

Fig. \ref{oldcrystals}b reports the normalized low-temperature
conductance of the same contact (top) and that of a different
contact with a higher $R_N$ =18.8 $\Omega$ (bottom). The
experimental data (symbols) are compared against the relevant 2D-BTK
fit (lines). It is worth noting that these curves (and \emph{all}
the curves obtained in crystals with reduced $T_c$) show a small
Andreev signal, whose amplitude is similar to that observed in
CeCoIn$_5$ \cite{park05,park08}. It seems however logical, in this
case, to ascribe the small signal to the greater amount of disorder
in the samples (and thus also on the surface) with respect to the
purest crystals, rather than to intrinsic phenomena as proposed in
\cite{fogelstrom10}. The small amplitude and the consequent absence
of very clear structures result in a greater uncertainty on the
fitting parameters, since different sets of parameters can give
equally good fits. Moreover, the fit requires a large $\Gamma$
(although still smaller than $\Delta$). As an alternative, and
according to \cite{fogelstrom10}, one could assume that a fraction
of the injected current tunnel into a non-superconducting band or
set of states, and does not contribute to Andreev reflection. In the
latter case, the values of $\Gamma$ would be smaller but an
additional parameter would be necessary.

Fig. \ref{oldcrystals}c shows an example of temperature dependence
of the gap extracted from the fit (symbols). The line is the curve
calculated within the Eliashberg theory, by using the \emph{same}
spectral function and coupling constant as for the sample with
T$_{c}$ = 18.5 K. The different $T_c$ and the different gap
amplitude arise only from the larger value of the quasiparticle
scattering rate $\gamma_{d}$ included in the Eliashberg equations to
account for the disorder. Incidentally, while $T_c=18.5$ K is
obtained by using $\gamma_{d}=0.25$ meV, the reduced $T_c=14.5$ K of
this particular contact requires $\gamma_{d}=1.6$ meV. Note that
also the gap ratio $2\Delta/k_B T_c=6.2 \pm 0.4$ is compatible with
that obtained in the purest crystals.

%\bibliographystyle{naturemag}
%\bibliography{MSWEBpublications}

\newpage

%\pagebreak

%\pagebreak

%\pagebreak

\end{document}